\documentclass[conference]{IEEEtran}
\IEEEoverridecommandlockouts
\usepackage{cite}
\usepackage{amsmath,amssymb,amsfonts}
\usepackage{graphicx}
\usepackage{textcomp}
\usepackage{xcolor}

\usepackage{amsthm}
\newtheorem{definition}{Definition}
\newtheorem{lemma}{Lemma}
\newtheorem{theorem}{Theorem}

\usepackage[utf8]{inputenc}

\usepackage{graphicx}

\usepackage{amsmath}
\usepackage{pifont}

\usepackage{xfrac}

\usepackage{etoolbox}

\usepackage{algorithm}
\usepackage{algorithmicx}
\usepackage{algpseudocode}
\usepackage{mdframed}

\usepackage{caption}
\usepackage{subcaption}
\usepackage[percent]{overpic}

\usepackage{xspace}
\usepackage{numprint}
\usepackage{url}

\usepackage{breakurl}
\usepackage{hyperref}
\usepackage{xcolor}
\usepackage{siunitx}
\usepackage{thmtools,thm-restate}

\usepackage{array}
\newcolumntype{x}[1]{>{\centering\arraybackslash\hspace{0pt}}p{#1}}

\usepackage{adjustbox}
\newcolumntype{R}[2]{>{\adjustbox{angle=#1,lap=\width-(#2)}\bgroup}l<{\egroup}}

\usepackage{wasysym}
\usepackage{bm} 

\newtoggle{draft}
\newtoggle{paper}
\newtoggle{preprint}
\newtoggle{impossibilityproof}
\newtoggle{correctnessproofinbody}

\togglefalse{draft}
\toggletrue{paper}
\toggletrue{preprint}

\iftoggle{paper}{
  \togglefalse{correctnessproofinbody}
  \togglefalse{impossibilityproof}

  \iftoggle{preprint}{
}{}
}{}

\iftoggle{preprint}{
  
\toggletrue{correctnessproofinbody}
  \toggletrue{impossibilityproof}
}{}

\newcommand{\commandtag}[1]{\textsc{\expandafter\MakeLowercase{#1}}\xspace}
\newcommand{\tc}{\textsf{\small TC}\xspace}
\newcommand{\tcdot}[1]{\textsf{\small TC.#1}}
\newcommand{\tcvdot}[2]{\textsf{\small TC$_{#1}$.#2}}

\newcommand{\request}{\commandtag{REQUEST}}
\newcommand{\reply}{\commandtag{REPLY}}

\newcommand{\checkpoint}{\commandtag{CHECKPOINT}}
\newcommand{\reqviewchange}{\commandtag{REQ-VIEW-CHANGE}}
\newcommand{\viewchange}{\commandtag{VIEW-CHANGE}}
\newcommand{\newview}{\commandtag{NEW-VIEW}}
\newcommand{\viewconfirm}{\commandtag{view-confirm}}
\newcommand{\orderrequest}{\commandtag{order-request}}

\newcommand{\remove}[1]{}

\newcommand{\largename}{\textsf{{SACZyzzyva}}\xspace}
\newcommand{\name}{\textsf{\small{\largename}}\xspace}
\newcommand{\smallname}{\textsf{\footnotesize{\largename}}\xspace}

\newcommand{\protocolstep}[3]{
  \vspace{\baselineskip}
  \par\noindent
  \newdimen\protocolstepwidth
  \protocolstepwidth=\linewidth
  \addtolength\protocolstepwidth{-2\fboxsep}
  \fbox{\begin{minipage}[c]{\protocolstepwidth}{{\textsf{\small#1}}}\end{minipage}}
  \vspace{0.2em}

\noindent \underline{Explanation:} \hangindent1em #2\par
\vspace{0.2em}
\ifstrempty{#3}{}{
\noindent \underline{Details:} \hangindent1em #3
}}

\newcommand{\pk}{\textsf{pk}\xspace}

\title{Making Speculative BFT Resilient \\ with Trusted Monotonic Counters}
\iftoggle{draft}{
\IEEEspecialpapernotice{(Draft; not for further distribution)}
}{}

\author{\\ \\ \\ \\}

\newcommand{\aaltoblock}[1]{\IEEEauthorblockA{
    Aalto University \\
    #1}}
\newcommand{\berkeleyblock}[1]{
  \IEEEauthorblockA{University of California, Berkeley \\
    #1}
}
\newcommand{\intelblock}[1]{
  \IEEEauthorblockA{Intel Labs\\
   #1}
}

\author{
  \IEEEauthorblockN{Lachlan J.\ Gunn}\aaltoblock{lachlan@gunn.ee}
  \and
  \IEEEauthorblockN{Jian Liu}\berkeleyblock{jian.liu@berkeley.edu}
  \and
  \IEEEauthorblockN{Bruno Vavala}\intelblock{bruno.vavala@intel.com}
  \and
  \IEEEauthorblockN{N.\ Asokan}\aaltoblock{asokan@acm.org}
}

\begin{document}

\maketitle

\thispagestyle{plain}
\pagestyle{plain}

\begin{abstract}
Consensus mechanisms used by popular distributed ledgers are highly scalable but
notoriously inefficient. Byzantine fault tolerance (BFT) protocols are efficient
but far less scalable. Speculative BFT protocols such as Zyzzyva and Zyzzyva5
are efficient and scalable but require a trade-off: Zyzzyva requires only $3f+1$
replicas to tolerate $f$ faults, but \emph{even a single slow replica} will make
Zyzzyva fall back to more expensive non-speculative operation. Zyzzyva5 does not
require a non-speculative fallback, but requires $5f+1$ replicas in order to
tolerate $f$ faults. BFT variants using hardware-assisted trusted components can
tolerate a greater proportion of faults, but require that \emph{every} replica
have this hardware.

We present \name, addressing these concerns: resilience to slow
replicas and requiring only $3f+1$ replicas, with only one replica needing an active monotonic counter at any given time.

We experimentally evaluate our protocols, demonstrating low latency and high scalability.
We prove that \name is optimally robust and that trusted components cannot
increase fault tolerance unless they are present in greater than two-thirds of
replicas.
 \end{abstract}

\section{Introduction}

Distributed ledger technology~\cite{HLSawtooth,androulaki2018hyperledger,
  bonneau2015sok} and cryptocurrencies~\cite{bitcoin,ethereum} have become the
great motivators for distributed consensus protocols today.
These applications demand scalability and performance over high-latency networks
such as the Internet.
Current approaches range from proof-of-work~\cite{bitcoin,ethereum} to Byzantine
fault tolerance (BFT)~\cite{lamport1982byzantine, barborak1993consensus, martin2005fast, pbft, minbft, gueta2018sbft, malkhi2018blockchain}.

Both approaches have significant drawbacks.
Proof of work derives its Sybil-resistance from the magnitude of its power consumption~\cite{bonneau2015sok}.
Furthermore, its scalability comes at the cost of eschewing transaction
finality~\cite{vukolic2015quest, vukolic2016eventually}. 
Conversely, BFT protocols~\cite{lamport1982byzantine} are computationally efficient,
but scale poorly. As traditionally formulated, these require two phases~\cite{lamport2006lower} and a quadratic number of messages~\cite{DolevReischuk}. However, a wide variety of improvements can be
obtained over classical results~\cite{pbft} by varying cryptographic~\cite{cosi},
failure-mode~\cite{minbft,fastbft},
timing~\cite{correia2010asynchronous,honeybadger}, and safety~\cite{Zyzzyva}
assumptions.

Zyzzyva~\cite{Zyzzyva,abraham2018revisiting} is the
simplest and most compelling of the BFT protocols. It takes a
\emph{speculative} approach that optimizes for the common case where no replicas are
faulty.   MinZyzzyva~\cite{minbft} improves on Zyzzyva by assuming that each
replica contains a trusted monotonic counter, whose integrity is guaranteed by hardware.
In particular, it reduces the total number of replicas needed to tolerate $f$
faults from $3f+1$ to $2f+1$.
\begin{table}
  \centering
  \def\arraystretch{1.1}
  \begin{tabular}{|c|c|c|c|c|}\multicolumn{1}{c}{} & \multicolumn{1}{c}{\textbf{Total \# of}} & \multicolumn{1}{c}{} & \multicolumn{2}{c}{\textbf{Monotonic counters}}  \\
    \multicolumn{1}{c}{\textbf{Protocol}} & \multicolumn{1}{c}{\textbf{Replicas}} & \multicolumn{1}{c}{\textbf{Resilience}} & \multicolumn{1}{x{1cm}}{\emph{Total}} & \multicolumn{1}{x{1cm}}{\emph{Active}} \\
    \hline
    Zyzzyva & $3f+1$ & \textcolor{red}{0} & \textbf{\textcolor{green!70!black}{-}} & \textbf{\textcolor{green!70!black}{-}} \\
    Zyzzyva5 & \textcolor{red}{$5f+1$} & \textcolor{green!70!black}{\underline{$\bm{f}$}}& \textbf{\textcolor{green!70!black}{-}} & \textbf{\textcolor{green!70!black}{-}} \\
    MinZyzzyva & \textcolor{green!70!black}{\underline{$\bm{2f+1}$}} & \textcolor{red}{0} & {\textcolor{red}{$2f+1$}} &  {\textcolor{red}{$2f+1$}} \\
    \textbf{{\smallname}} & $3f+1$ & \textcolor{green!70!black}{\underline{$\bm{f}$}} & \textcolor{green!70!black}{\underline{$\bm{f+1}$}} & \textbf{\textcolor{green!70!black}{\underline{1}}} \\
    \hline
  \end{tabular}
  \caption{Comparison of speculative BFT protocols tolerating $f$ faults.
    Resilience refers to the maximum number of replicas that can be non-responsive
    without falling back to non-speculative operation.}
  \label{tbl:comparison}
\end{table}

This assumption, that every replica is equipped with a
trusted component, is often unrealistic. In the real world,
only some devices will have the necessary hardware, especially when new hardware
is being rolled out.  Even if eventually all replicas have the necessary hardware
support, over time some hardware platforms will become obsolete either because
have become outdated in comparison to newly-released hardware, or because trust
in them has been revoked in response to some vulnerability.
Protocols that require trusted components in every
participant are thus fragile.

Speculative BFT protocols have extremely simple and efficient speculative
execution paths when there are no faults or delays. In the event of a fault,
Zyzzyva and MinZyzzyva have the client execute a non-speculative fallback
sacrificing performance. This results in a major drawback: if even a single
replica fails to respond to the client, the protocols immediately fall back to
non-speculative execution, unlike non-speculative protocols which concern
themselves mainly with faulty \emph{primaries}~\cite{CWADM09}. Realistic
communication networks like the Internet are only partially synchronous. In such
networks, a single \emph{slow}---not necessarily faulty---replica can trigger the
non-speculative execution for each protocol run, thereby undermining the
efficiency promise of the speculative approach. Speculative variants like
Zyzzyva5~\cite{Zyzzyva} minimize the need for non-speculative fallback, but have
lower fault-tolerance, requiring $5f+1$ replicas to tolerate $f$ faults.

In this paper, we present Single Active Counter Zyzzyva (\name), which overcomes these drawbacks. It requires
only a single replica, the primary, to have an active monotonic counter,
and eliminates the need
for a non-speculative fallback (as in Zyzzyva5), thus allowing \name to tolerate a subset of
replicas being slow, while requiring only $3f+1$ replicas (as in Zyzzyva).  We
compare \name to other speculative BFT protocols in Table~\ref{tbl:comparison}. The same principles that we use in \name can
be applied in other settings: other BFT protocols can be adapted to use our \emph{single
  active counter approach}, resulting in lower latency while avoiding the need
to equip all replicas with hardware-supported monotonic counters.

The cost of supporting this heterogeneity---of not requiring that all replicas
have trusted components---is the need for more replicas to tolerate the same
number of faults $f$: $3f+1$ in \name{}, compared to $2f+1$ in MinZyzzyva. We
show that \name{} is optimally fault tolerant. Specifically, it is not possible
to tolerate more than $\lfloor (n-1)/3 \rfloor$ failures---as \name does---unless more than two thirds of
parties have access to a trusted component (as MinZyzzyva does).

In summary, our main contributions are as follows:

\begin{itemize}
    \item We propose $\name$ (Section~\ref{sec:saczyzzyva}), a Zyzzyva variant
      that tolerates $\lfloor (n-1)/3 \rfloor$ faults and uses a trusted monotonic counter to eliminate the need for a
      non-speculative fallback, making it more robust to slow replicas.
    \item We implement and evaluate \name over both
      low- and high-latency networks (Section~\ref{sec:evaluation}), showing that \name
      transaction latencies increase at a rate of less than \SI{40}{\micro\second} per
      additional replica.
    \item We show that the use of trusted components in a consensus protocol
      cannot increase fault-tolerance unless more than two thirds of parties
      have a trusted component (Section~\ref{sec:impossibility}).
\end{itemize}

 \section{Preliminaries}

\subsection{Zyzzyva}
Zyzzyva~\cite{Zyzzyva} is an efficient Byzantine-fault-tolerant state-machine replication protocol which uses \emph{speculation} to
reduce the replication overhead, at the cost of requiring rollback in some instances.
The replicas receive requests ordered by the primary, and immediately reply to clients without running an expensive consensus protocol.
Based on the received replies, clients are able to detect inconsistencies and can help the replicas achieve a consistent state.
In fault-free executions with network delays that do not trigger protocol timeouts, no further action is required by clients, thereby making the protocol simple and fast.

The protocol works as follows:
The client sends a request to the primary, which in turn proposes an order and forwards it to the other replicas.
The replicas speculate that the primary's proposal is consistent and reply to the client.
If the client receives matching replies from all replicas, then
speculative execution is successful and the request is guaranteed persistent.
However, if after some timeout $T_\mathrm{client}$ the client receives between
$2f+1$ and $3f$ matching replies, the
client executes a non-speculative fallback: it broadcasts the responses that it
has received to all replicas, and waits for $2f+1$ acknowledgements. The
replicas acknowledge the commit certificate if it is consistent with the local
history of ordered requests.  This non-speculative fallback allows for operation
in the presence of faults, but comes with significant latency costs.

Finally, if acknowledgements are not received, the client broadcasts the
request to all replicas, who communicate with the primary to assign a sequence number and execute it.

Zyzzyva is efficient and scalable, but this efficiency comes at a
price, in the form of fragility. If even a single
replica is faulty, or network conditions cause a single message to be delayed
beyond the timeouts, speculative execution fails and the client must execute its
non-speculative fallback, requiring at least two additional rounds of
communication, in addition to the time spent waiting for the timeout.  This
negates Zyzzyva's main contribution---its high performance---especially over the internet where Zyzzyva's
small communication footprint would otherwise be most useful.

A variant, Zyzzyva5~\cite[Section~4.1]{Zyzzyva}, was introduced along with
Zyzzyva, which avoided this non-speculative fallback at the cost of
fault-tolerance by increasing the number of replicas from $3f+1$ to $5f+1$, and
allowing requests to complete after $4f+1$ responses.  With these thresholds,
all requests complete speculatively, but at the cost of Zyzzyva5
only tolerating $\lfloor (n-1)/5 \rfloor$ faults in comparison to Zyzzyva's $\lfloor (n-1)/3 \rfloor$.

\subsection{Hybridization and trusted components}
Another way to improve on classical BFT results is to use
\emph{hybridization}~\cite{hybridization}, in which replicas contain several
components of different failure modes. Under this model, failed replicas cannot
behave completely arbitrarily; instead, they are limited by their non-Byzantine
components.

A common approach is to design the replicas around a \emph{trusted component},
whose output can be authenticated by other parties and is subject only to
crash-failures.  This can be achieved with the aid of the
hardware-assisted trusted execution environments (TEEs) that exist in many modern CPUs.

TEEs protect the execution of a security-critical piece of application from
potentially-compromised applications, system administrators and the operating
system itself. By the process of \emph{remote attestation}, they can securely
communicate the existence of such trusted components to an external verifier,
allowing other parties to rely on the security guarantees provided by the
hardware. Examples of such hardware mechanisms are Intel SGX~\cite{sgx} and ARM
TrustZone~\cite{trustzone}.

TEEs are highly general; in concrete protocols, we generally do not consider
their full functionality, but instead use them to implement
more limited trusted functionality that can be effectively reasoned about.  An
especially popular such functionality is the \emph{trusted monotonic counter}.

A trusted monotonic counter uses these hardware security features to realize a
verifiably monotonically increasing counter.

Let $\langle M \rangle_{X}$ indicate that a message $M$ has been signed by some entity
$X$. A trusted monotonic counter component $\tc$ is assumed to have a
well-known public key---for example, established with remote attestation---and provide the following
interface:
\begin{itemize}
  \item $\tc_\textrm{inst}, \langle \textsf{pk}_{\tc_\textrm{inst}} \rangle_{\tc} \gets
    \tcdot{Init}()$: Create a new trusted monotonic
    counter instance $\tc_\textrm{inst}$, with initial state $c = 0$ and public key $\textsf{pk}_{\tc_\textrm{inst}}$.
  \item $\langle c, m \rangle_{{\tc_\mathrm{inst}}} \leftarrow \tcvdot{\textrm{inst}}{Increment}(m)$: Update
    the counter state $c \leftarrow c + 1$, returning a signed tuple linking a message
    $m$ to this particular increment operation and trusted monotonic counter
    instance.
\end{itemize}

Trusted monotonic counters are used in BFT protocols such as MinBFT~\cite{minbft} to prevent message
equivocation. A trusted monotonic counter value can be attached to a message in order to
detect whether the sender communicated the same data to all recipients. If the
sender equivocates, different messages will have different counter values, this
being detectable as a `hole' in the set of counter
values~\cite{levin2009trinc,matetic2017rote,strackx2016ariadne,minbft}.
Persistent hardware-backed versions of such counters are available within
TPMs~\cite{TCGTPM} and the Intel SGX~\cite{sgx} platform; alternatively, a TEE can
be used to implement a memory-backed monotonic counters that offers high
performance at the cost of ephemerality or replication~\cite{matetic2017rote}.

\section{Model and Problem Statement}

\subsection{Network Model}

In this paper we consider the weak-synchrony model~\cite{pbft,honeybadger}.
Messages and computation can be arbitrarily
delayed, but the delay $\mathrm{delay}(t)$ of a message sent at time $t$ cannot
grow faster than the timeout period---which may vary adaptively---indefinitely.

This model permits polynomially-increasing delays when exponential backoff is used to
increase message timeouts~\cite[§3.1]{honeybadger}. However, it does not allow an
adversary to continually delay messages so that they arrive after the
exponentially-increasing timeouts, thereby achieving eventual synchrony~\cite{Zyzzyva}.
This model enables us to analyze liveness during a period of synchrony that will
eventually occur. This also avoids the well-known FLP impossibility result~\cite{fischer1985impossibility}, which showed that it is not possible to achieve consensus in a fully asynchronous system~\cite{lynch-distributed-algorithms}.

\subsection{System Model}\label{sec:system-model}
We consider a distributed system of $n$ replicas, of which up to $f$ may be faulty.
We suppose that \emph{some}, but not \emph{all}, replicas are equipped 
with a trusted component and, in particular, with a trusted monotonic counter. The result
is that $b$ out of $n$ replicas can, if faulty, behave completely arbitrarily,
whereas the other $n-b$ replicas, if they fail, are assumed to be limited in
their behavior by the trusted component.

\subsection{Problem Statement}
Our goal is to build an efficient state-machine replication protocol that allows the replicas to complete a request
\begin{itemize}
\item in a linear (in $n$) number of messages, and
\item without significant performance reductions in the event of up to $f$
  faults.
\end{itemize} 

We borrow from Zyzzyva~\cite{Zyzzyva} the properties that our BFT protocol must
satisfy to be correct. The first one is \emph{safety}: suppose that from the
perspective of some client, a request completes
with a response indicating a history $H$---a sequence of ordered and completed
requests---then the history of any other completed request as seen by any other
client is a prefix of $H$, or
vice-versa. Hence, from the perspective of the client, the state machine history
never diverges, even if that of individual replicas might.
The second one is \emph{liveness}: any request issued by a correct client
eventually completes.
 
\section{SACZyzzyva}
\label{chap:design}
\label{sec:saczyzzyva}

In the original Zyzzyva with $3f+1$ replicas, a request is included in a new
  view only when it appears in $f+1$ out of $2f+1$ \viewchange messages. Since
  up to $f$ of these \viewchange messages may be from faulty replicas, this
  means that \emph{every} correct replica must execute the request in order to
  guarantee that a speculatively-executed request will be included in the
  history of future views.

  The MinZyzzyva~\cite{minbft} protocol uses a trusted monotonic counter in each
  replica to order requests and prevent equivocation. In doing so, it reduces to
  $2f+1$ the number of replicas needed to tolerate $f$ faults, but does not
  change the protocol in a fundamental way. However, the MinZyzzyva view-change
  protocol differs from that of Zyzzyva, with the initial state of a view being
  determined as in MinBFT~\cite[p.~8]{minbft}: a request is included in the
  history of a new view when it appears in \emph{any} \viewchange message. This
  means that MinZyzzyva needs only one copy of a request to appear in
  any set of $f+1$ view-change messages in order to guarantee that
  speculatively-executed requests are not lost. By modifying Zyzzyva to order
  requests within a view using a trusted monotonic counter in the primary, we
  can use the same inclusion criteria during view-changes as in the MinBFT
  protocols, allowing requests to safely complete after only $2f+1$ responses,
  eliminating the need for a non-speculative fallback.  We dub this protocol
  \emph{Single Active Counter Zyzzyva} (\name).

The basic principle of $\name$ is to use a trusted monotonic counter in the primary to bind
a sequence of consecutive counter values to incoming requests, ordering requests while avoiding the need for communication between
replicas, whether directly or via the client.
It does this by signing a tuple
consisting of the cryptographic hash of the request and a fresh (i.e. has not been used before) counter value. 
This is then sent to all replicas in an
\orderrequest message. Because the primary is the only replica that actively maintains
a counter, we call this counter the ``Single Active Counter'' (SAC) construct.  We
therefore require only that $f+1$ replicas have a trusted component, enough that
there will always be at least one correct replica that can function as primary.

\begin{figure}[!h]
\centering
  \includegraphics[width=\linewidth]{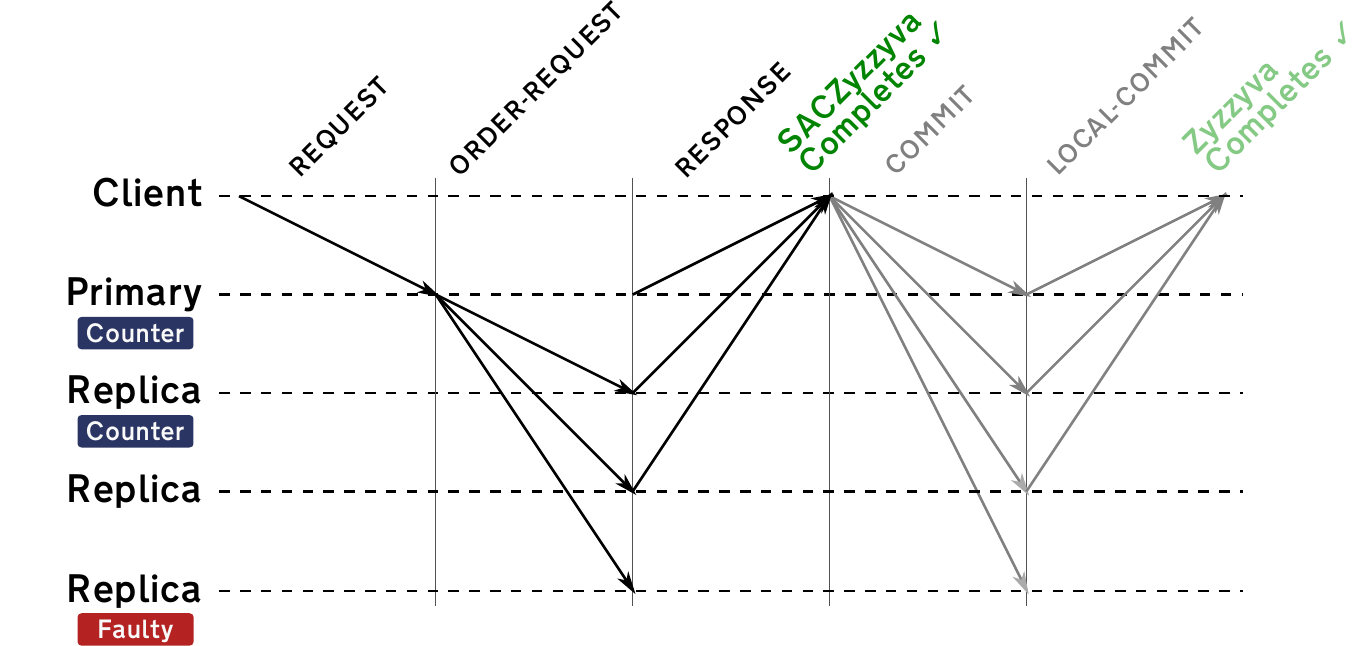}
\caption{The communication patterns of Zyzzyva and \name with one faulty
  replica.  Without faults or network
  delays, Zyzzyva and \name have identical communication patterns, but if any
  replicas are faulty, as illustrated, Zyzzyva requires two extra rounds of
  communication, shown in gray.}
  \label{fig:sacZyzzyva_comm}
\end{figure}

Figure~\ref{fig:sacZyzzyva_comm} shows the communication pattern of \name.
As in the original Zyzzyva, the primary gathers the requests from clients and sends them to all replicas in a \orderrequest message. 
The main difference is that the \orderrequest message is bound to a monotonic counter
value to prevent equivocation by the primary.
All replicas execute the requests and reply to the client directly if the
trusted monotonic counter value is sequential to those that the primary has previously sent.
If the client receives $2f+1$ replies with matching values and histories, it considers the request complete.
Otherwise, it repeatedly sends the requests directly to the replicas, so that
they can detect misbehavior by the primary and so elect a new one.

The protocol is described below in greater detail. The basic steps are shown in
boxes below; further explanation and specifics appear beneath each step.  We
assume that there is some well-known mapping from view-numbers to
primary replicas. One such mapping is $p_v = v \ \mathrm{mod}\ n_\textsc{tmc}$, where the replicas are numbered such that the
first $n_\textsc{tmc} > f$ replicas possess a trusted monotonic counter.  For
simplicity of the protocol description, when a replica broadcasts a message to all
replicas, this includes itself.

The correct replicas are assumed to
know each others' identities and primary keys when the system is initialized, as
well as the public key of the first primary's monotonic counter instance.

\subsection{Agreement protocol}
Requests are initiated similarly to the original Zyzzyva. However, unlike with
Zyzzyva, only $2f+1$ replies are needed before an operation is accepted as
complete. After receiving a request from the client, the primary binds a counter
value to the request and then sends it to the replicas for execution, who reply
directly to the client.

\protocolstep{
  C-1. The client sends a request to the primary.
}{
Since all requests must pass through the primary in order to be executed,
the client can initially send the request to the primary only.
}{
  The client sends a message
  \[m_\mathrm{request} = \langle \request, \mathrm{op},
  \mathrm{addr}_\mathrm{client},  \mathrm{id}_\mathrm{client} \rangle_c\] to the primary,
  where $\mathrm{op}$ is the requested operation,
  $\mathrm{addr}_\mathrm{client}$ is the address of the client to which the
  replicas must reply, and $\mathrm{id}_\mathrm{client}$ is a monotonically
  increasing identifier used to identify whether a request has already been
  executed.
}

\protocolstep{  
  R-1. Upon receiving a valid \request message, \emph{the
  primary} binds the request to a counter value and then broadcasts it to all
  replicas as an \orderrequest message.
}{
  Note that the primary only
  needs to act on \emph{valid} requests;
  a request might be invalid if it is not syntactically
  correct, but there may be other cases, such as if the state-machine being
  replicated includes client authentication or replay protection functionality.
}{
  After receiving a request
  \[m_\mathrm{request} = \langle \request, \mathrm{op},
                          \mathrm{addr}_\mathrm{client},
                          \mathrm{id}_\mathrm{client} \rangle_c, \]
  the primary verifies that $m_\mathrm{request}$ is valid and then binds a
  request number to it, using its trusted monotonic counter to obtain an
  `ordering certificate'
  \[
    C_\mathrm{order} \leftarrow \tcvdot{v}{Increment}(H(m_\mathrm{request})),
  \]
  which it includes in the \orderrequest message 
  \[
    m_\mathrm{order-request} = \langle \orderrequest, v,
    C_\mathrm{order}, m_\mathrm{request}\rangle_p
  \]
  that is broadcast to all replicas.
}

\protocolstep{
  R-2. Upon receiving an \orderrequest message for the current view with a
  counter value greater than that of the last executed request, each
  replica executes the request contained in the message and responds directly to
  the client, obtaining any previous requests needed by sending \commandtag{fill-hole}
  messages to the primary.
}{
  \name replicas execute requests immediately.  However,
  since the primary uses its trusted monotonic counter to number the requests
  within each view, the agreement protocol ends here and no \emph{commit certificate}
  subprotocol is needed as in Zyzzyva~\cite[Section 3.2, Steps 4.b.*]{Zyzzyva}.
}{
Replicas execute requests immediately if
they have executed all previous requests, and store the largest identifier
of each executed request from each client. If any previous requests have not
yet been executed, the replica demands a copy of the corresponding \orderrequest messages from the
primary by sending messages $\langle \commandtag{fill-hole}, v, i \rangle$.  If the replica does not receive a response within time $T$, the primary is deemed to be faulty and so the replica requests
a view-change.
}

\protocolstep{
  R-2a. Upon receiving a $\langle\commandtag{fill-hole}, v, i\rangle$ message
  from another replica, a replica responds with the \orderrequest message for request
  $i$ from view $v$ if it is in its history.
}{
 Since not every replica is guaranteed to see every request, they need some way
 to ``catch up'' when there is a hole in their execution history.  Correct
 replicas are guaranteed to receive a response from a correct primary during periods of synchrony, as
 the sender will only send this message after receiving an \orderrequest message for a
 later message in the same view, and so the necessary \orderrequest message will
 always be in the history.
}{}

\protocolstep{
  C-2a. The client waits for $2f+1$ matching replies from distinct replicas;
  it then accepts the response contained in these replies.
}{
During periods of synchrony, when the primary is correct and at least $2f+1$
replicas are correct overall, the client will receive sufficient reply
messages to accept a response.  This is in contrast to Zyzzyva, which can only
accept at this point only after receiving responses from all $3f+1$ replicas,
thus necessitating additional steps as a fallback.
}{
  The client receives $2f+1$ messages \[\langle \reply, \langle m_\mathrm{order-request} \rangle_{p_v}, \mathrm{response}
  \rangle_{i}\] from distinct replicas $\{i\}$ for some valid \orderrequest message
$m_\mathrm{order-request}$ in view $v$ and response ${\mathrm{response}}$.  The client then
  accepts the value ${\mathrm{response}}$ as the response to the request
  contained in $m_\mathrm{order-request}$.
}

\protocolstep{
  C-2b. After each time interval $T_\mathrm{client}$ that the client has not received $2f+1$
  matching replies from distinct replicas,
  the client broadcasts the request to all replicas.
}{
If the client does not receive a timely quorum of responses, then it is possible
that the replicas did not all receive the request from the primary.  In this
case, the client sends the request to the replicas directly, so that they can
determine whether the primary is willing to order the request, and initiate a
view-change if not.
}{
  The client broadcasts to all replicas the message $m_\mathrm{request}$,
  previously sent to the primary in step C-1.
}

\protocolstep{
  R-3. Upon receiving a \commandtag{request} message whose
  $\mathrm{id}_\text{client}$ is greater than the last cached identifier for
  that client, a replica will send it to the primary, and then wait for time $T$
  to receive a \orderrequest message that will be processed as in
  step R-2, otherwise requesting a view-change and broadcasting the request to
  all replicas.
}{
  Routing requests through the primary makes it into a single-point-of-failure.
  In order to prevent the primary from dropping requests---and thus violating
  liveness---the client rebroadcasts its request to the replicas so that they
  can submit the request on the client's behalf, giving the replicas the
  opportunity to observe the primary's misbehavior first-hand and then trigger
  a view-change.  As a side-effect, this also allows request processing to
  continue when the client does not know the current primary.
}{
  In addition to the above, a replica receiving a \request from another replica
  responds with the corresponding \orderrequest if it has it.
}

\subsection{View-change protocol}\label{sec:saczyzzyva-vc}
In the Zyzzyva protocol, a request is included in the history of a new view if
and only if there are $f+1$ \viewchange messages available containing the
request. As there might be only $f+1$ \viewchange messages from correct
replicas, to be certain that a request will be included in any new view, the
client therefore needs to ensure that \emph{every} correct replica has
responded. In the SACZyzzyva view-change protocol, the canonical ordering
provided by the trusted monotonic counter allows us to safely include requests
whose ordering exists in even a single \viewchange message. The client therefore
needs only $2f+1$ replies in order to be certain that a request will persist
across the next view-change.

\protocolstep{
  VC-1. When a replica requests a view-change, it broadcasts a \reqviewchange
  message to all other replicas and increases its timeout $T$ in some
  implementation-defined way.
}{
  This part of the view-change protocol remains unchanged from Zyzzyva.
}{The
  replica that has witnessed misbehavior of the primary of view $v$
  broadcasts a message $\langle \reqviewchange, v \rangle_i$ to all replicas.
}

\protocolstep{
  VC-2. Upon receiving $f+1$ \reqviewchange messages for the current view $v$, a replica stops
  processing requests in the current view and broadcasts a \viewchange message
  to all replicas.  If the view-change does not complete within
  time $T$, the replica requests a new view-change.
}{
  Since there is no \emph{prima-facie} evidence of misbehavior by
  the primary, before committing to a view-change each replica waits until misbehavior has
  been reported by at least $f+1$ replicas, so that it can prove to others with
  its \viewchange message that at least one report is genuine.
}{
  More specifically, replica $i$ sets its current view-number to $v+1$ and broadcasts a message
  $\langle\textsc{view-change}, v+1, i, V, R, \{r_i\} \rangle_{i}$, where $v+1$ is the new
  view-number, $V$ is the most recent view or checkpoint certificate, $\{r_i\}$ is the set
  of requests that it has executed in view $v$, and $R$ is a set of $f+1$
  \reqviewchange messages requests for view $v$.
}

\protocolstep{
  VC-3. Upon receiving $2f+1$ \viewchange messages for
  a new view $v$, the primary
  for $v$ instantiates a trusted monotonic counter instance and broadcasts a \newview
  message to the replicas.
}{
  With $2f+1$ \viewchange messages, any request that has been accepted by a
  correct client in the last view must be present in at least one of them.  This
  means that the primary can now safely propose a new view.  Rather than
  directly including the view's initial state, the \newview message includes the
  $2f+1$ \viewchange messages directly, so that the other replicas can
  themselves verify that all completed requests are included in the history of
  the new view.
}{
  If the view-number in these messages is less than that of this replica's
  current view number, then this 
  step is ignored.  Otherwise, the new primary runs $\tcdot{Init}()$, yielding a new trusted monotonic
  counter $\tc_v$ with corresponding public key $\pk_{\tc_v}$,
  then  broadcasts the message $\langle \newview, \langle \pk_{\tc_{v}} \rangle_\tc,
  \{m_{\textsc{vc},i}\}\rangle_{p_v}$ to all replicas, where
  $\{m_{\textsc{vc},i}\}$ are the $2f+1$ valid view-change messages that has
  been received.
}

\protocolstep{
  VC-4.  Upon receiving the first valid \newview message for view $v$, each replica
  broadcasts a \viewconfirm message containing a hash of the \newview message it
  has received.
}{
  Though a valid \newview message is guaranteed to contain every completed
  request, a faulty primary can provide a different set of \viewchange messages
  to each replica, causing them to disagree on whether uncompleted requests are
  included.  This step ensures that all completed requests will build on
  the same \newview message.
}{
  For the \newview message to be valid, it must contain \viewchange messages
  from $2f+1$ distinct replicas, and a public key that has been verified to
  belong to a trusted monotonic counter instance. If the view-number in the
  \newview message is less than that of this replica's
  current view, then this message can be ignored.  Otherwise,  
  after receiving a \newview message $m$ for view $v$ for the first time, each
  replica broadcasts a message $\langle \viewconfirm, v, i, H(m) \rangle_i$ to all replicas.
}

\protocolstep{
  VC-5.  Upon receiving $2f+1$ matching \viewconfirm messages from distinct
  replicas confirming the \newview message from step VC-4, the view-change
  completes, and each replica begins
  to process requests in the new view.
}{
  After receiving $2f+1$ matching \viewconfirm messages, a correct replica 
  can be certain that no other correct replica will process requests
  in this view with a different starting state.
}{
  Consistency in this case means that all $2f+1$ messages have identical
  view-numbers $v$ and \newview hashes $H(m)$.
  The starting state for this view is taken to be that of highest-numbered view
  with a certificate in any of the \viewchange messages in the confirmed
  \newview message, extended with the longest consecutive sequence of requests
  in any of the same \viewchange messages containing this view.  Putting a
  replica into this state may require rolling-back some previously-executed
  requests, and making it necessary to maintain enough information to roll back
  to the last checkpoint, or in extreme cases to carry out state transfer as
  in~\cite{pbft}.  These $2f+1$ \viewconfirm messages are stored as a
  \emph{view certificate}.
}

\subsection{Checkpointing Protocol}
Since it is possible that a view-change might require a replica to roll-back
some already-executed requests in the latest view, replicas must maintain enough
information to rewind their state to the last confirmed transaction.   To keep the required storage from growing
without bound, Zyzzyva includes a
checkpoint protocol~\cite[Section~3.1]{Zyzzyva} taken from that of
PBFT~\cite[Section~4.3]{pbft}; we do not reproduce all of the details, but
sketch it here.
\protocolstep{
  CP-1 (sketch). Every $N$ requests, each replica broadcasts a
  \checkpoint message containing the current view certificate, the most
  recently-executed request number, and a hash of the current state to all
  replicas.
}{
  Since a correct replica will include in its \viewchange messages every request
  that it has executed, a \checkpoint message is a commitment to include
  all of these requests in future \viewchange messages.
}{}

\protocolstep{
  CP-2 (sketch).  After receiving $2f+1$ matching \checkpoint messages for the
  current view, a replica considers the \checkpoint to be \emph{stable}, and
  discards all \orderrequest messages from before the checkpoint.
}{
  Once $2f+1$ replicas have commit to including a request
  in their future \viewchange messages, then it is guaranteed that at least one
  correct replica from among them will have their \viewchange message appear in
  any future successful view-change.  The \checkpoint messages are stored as the
  latest \emph{checkpoint certificate}.
}{}

\noindent In this sketch we do not include e.g.\ low- and high-water marks; full details
can be found in~\cite[Section~4.3]{pbft}.

\iftoggle{correctnessproofinbody}{
\iftoggle{correctnessproofinbody}{
  \subsection{Correctness}\label{sec:correctness}
}{
\section{Correctness}\label{sec:correctness}
}

The safety and liveness properties of \name are defined from the point
of view of the client: the states of the replicas may diverge, so long as
the histories returned with completed requests do not diverge.

We recall that \name uses $n = 3f+1$ replicas in order to tolerate $f$ faults,
with $n_\textsc{tmc} > f$ replicas having a trusted monotonic counter. We are
therefore guaranteed that any set of $f+1$ replicas will always contain at least
one correct replica, and that any two sets of $2f+1$ replicas will always
contain at least one correct replica in their intersection.

We suppose as well that there exists some well-known mapping from view-numbers
to trusted monotonic counter-equipped replicas such that at least one correct
replica will be chosen infinitely many times. One suitable mapping is $p_v = v
\ \mathrm{mod}\ n_\textsc{tmc}$, where the replicas are numbered such that the
first $n_\textsc{tmc}$ replicas possess a trusted monotonic counter.

\subsubsection{Safety}
We show that the histories of completed requests can never diverge. A history
$H$ is a sequence of requests $m_0, m_1, \ldots$ that are executed in turn. 
We use the notation $A \sqsubseteq B$ to indicate that $A$ is a non-strict prefix of $B$.

\begin{lemma}[Consistency of the initial view state]
  \label{lem:sacz-safety-history}
  Let $H_1$ and $H_2$ be the request histories and $\pk_1$ and $\pk_2$ the
  trusted-monotonic-counter public keys held any two correct replicas that
  execute any two requests in view $v$; these requests may or may not be
  distinct. Then, $\pk_1 = \pk_2$, and $H_1$ and $H_2$ are identical with
  respect to requests prior to view $v$.
  
  \begin{proof}
    If $v$ corresponds to the first view, then the history of prior views is
    empty in both cases, and the public keys form part of the initial state, and
    so the lemma is trivially true.

    Otherwise, each correct replica executing a request in view $v$ must have
    received at least $2f+1$ \viewconfirm message from distinct
    replicas for view $v$ containing the same hashed \newview message (Step
    VC-5). The $2f+1$ such messages received by each
    replica must have in common at least one correct sender. Each correct
    replica produces only a single \viewconfirm message---Step VC-4---so the
    consistent set of \viewconfirm messages must confirm the same \newview, and
    thus both replicas accept the same public keys and history as the initial state of the view.
  \end{proof}
\end{lemma}

\begin{lemma}[Histories of completed requests do not diverge within a single
  view]\label{lem:sacz-safety-intraview}
  Let requests $r_x$ and $r_y$ complete in view $v$, and let $H_x$ and $H_y$ be
  the request histories of any two correct replicas immediately after they
  executing $r_x$ and $r_y$ respectively. Then, one history is a prefix of the
  other---that is, either $H_x \sqsubseteq H_y$ or $H_y \sqsubseteq H_x$.

  \begin{proof}
    A correct replica responds only after having received messages $\langle \orderrequest,
    v, C_i, m_\mathrm{request}\rangle$ from the primary $p$ with sequential ordering
    certificates $C_i$ for every $m_\mathrm{request}$ in its history of this
    view (Step R-2). As $C_i$ can be obtained only by \tcdot{Increment} and includes
    $H(m_\mathrm{request})$, for any $C_i$ there is at most one request $m_i$
    for each $i$ such that any replica has received $\langle \orderrequest, v, C_i, m_i \rangle$,
    and therefore the histories of all correct replicas within view $v$ are
    identical except for partial truncation of a common suffix.  Hence, the
    history of any correct replica is either prefixed by or a prefix of any other.
  \end{proof}
\end{lemma}

\begin{lemma}[Completed requests are never omitted from history by a
  view-change]\label{lem:sacz-safety-vc-history}
  Let $H$ be the history of all completed requests up to and including
  view $v$. Then, for all
  views $v' > v$, a correct replica executing a request in view $v'$ includes $H$ in its
  history.
  \begin{proof}

    Let $v' > v$ have primary $p'$.
    By Lemma~\ref{lem:sacz-safety-history}, all correct replicas executing
    requests in view $v'$ will have identical histories for views prior to $v'$.
    
    We proceed by strong induction to show that this
    history is prefixed by $H$.

    \textbf{Base case.} Let $v' = v + 1$. For a correct replica to respond to a
    request in view $v'$, it must receive a \newview message containing $2f+1$
    \viewchange messages from distinct replicas. At least one of these
    \viewchange messages must be from a replica that is correct and has executed
    the last---and hence all prior---completed requests in $H$. Therefore $H$ will be a
    prefix of the history that this replica computes for view $v'$,
    and so $H$ will be in the history of any correct replica that
    begins executing requests in view $v'$.

    \textbf{Inductive case.} Let the supposition hold for all $v''$ such that $v
    < v'' < v'$.

    From step VC-5, the history of view $v'$ as confirmed by any correct replica
    is prefixed by the history of the most recent view $v_*$ for which a
    view-change certificate---or a checkpoint-certificate, which contains the
    corresponding view-change certificate---is available in one of the
    \commandtag{view-change} messages being confirmed, along with all subsequent
    requests in view $v_*$ for which an order-request message is available in one of
    the same \commandtag{view-change} messages.

    We will always have that $v_* \ge v$, as at least one of the $2f+1$
    \commandtag{view-change} messages must be from a correct node that executed
    $r$, and therefore has a view-change certificate for view $v$.

    If $v_* = v$, then the result is trivial: any set of
    $2f+1$ \commandtag{view-change} messages in a valid
    \commandtag{new-view} will include one from a correct node that executed
    the final request $r \in H$ in view $v$, and therefore a view-change certificate for view $v$ and the
    \orderrequest messages for $r$ and its predecessors are included.

    If $v_* > v$, then by supposition $r$ and its history are a prefix of the
    history of $v_*$, whose history is itself a prefix of the history of $v'$,
    which is what we wanted. \openbox

    The history of any correct replica that
    executes a request in view $v'$ is prefixed the computed history of view
    $v'$, and is therefore prefixed by $H$.
  \end{proof}
\end{lemma}

\begin{theorem}[Safety]
  Let requests $r_x$ and $r_y$ complete with histories $H_x$ and $H_y$ at any
  two replicas that have just executed requests $r_x$ and $r_y$ respectively. Then,
  one history is a prefix of the other---that is, either $H_x \sqsubseteq H_y$ or $H_y \sqsubseteq
  H_x$.
  \begin{proof}
    Suppose $r_x$ and $r_y$ complete in views $v_x$ and $v_y$ respectively. If
    $v_x = v_y$, then the theorem follows trivially from
    Lemmas~\ref{lem:sacz-safety-intraview}---for the part of the history in $v_x =
    v_y$---and~\ref{lem:sacz-safety-history}---for the history of earlier views.

    Otherwise, suppose without loss of generality that $v_x < v_y$. Then, by
    Lemma~\ref{lem:sacz-safety-vc-history}, the history of completed requests up
    to view $v_x$ is a prefix of $H_y$.  Since $r_x$ completes in view $v_x$, we
    therefore have that $H_x \sqsubseteq H_y$.
  \end{proof}
  
\end{theorem}

\subsubsection{Liveness}

We show that a request by a correct client eventually completes.
We say a view is {\em stable} if the primary is correct and enough time has
passed that network delays are less than the timeout period of the protocol.
The proof follows similarly to that of~\cite{fastbft}.

\begin{lemma}
\label{liveness-sac}
During a stable view, a request by a correct client will complete.

\begin{proof}
Since the primary is correct, a valid $\orderrequest$ message will be sent to
all replicas.  Since the network is in a period of synchrony,
the request will eventually complete, the client receiving at least $2f+1$
replies. 
\end{proof}
\end{lemma}

\begin{lemma}

\label{liveness2-sac}

For an unstable view $v$, either all requests will complete, or the view will
eventually change to a stable one.

\begin{proof}

  Suppose a client makes a request during an unstable view. Then, two things may
  happen: the primary provides a consistent ordering to $2f+1$ replicas that
  respond to the client before the client times out, in which case the request
  completes, or it does not.

  Suppose the client times out. Then, then all $2f+1$ correct replicas will
  eventually receive the request directly from the client (step C-2B), and those
  that have not already replied to the client will forward it to the primary
  (step R-3), setting a timeout.  If no correct replicas receive the
  corresponding \orderrequest, then all $2f+1$ of them will request a view
  change, leading to all correct replicas initiating a view-change.  Otherwise,
  if at least one correct replica receives the corresponding \orderrequest, then
  it will receive the requests forwarded by the other replicas in step R-3, and
  respond with the \orderrequest.  Thus all correct replicas will eventually
  receive the \orderrequest and respond to the client if they have not already
  begun a view-change.

  Therefore, either the request completes or all correct replicas eventually
  begin a view-change.

  If any correct replica commits to a view change, then there are three possible outcomes:

\begin{enumerate}

\item {\em All correct replicas change to a stable view.}
\item {\em All correct replicas change to an unstable view:} the client resends its request, which
  either completes or results in a further view-change (as above).

\item {\em At least one correct replica does not change view:}
  if any correct replica commits to a view-change, eventually so will all
  others.
  If at least $f+1$ correct replicas do not receive confirmation of the
  new view before timing out, then a further view-change will occur.  Otherwise,
  when the client resends its request, it will either complete or result in a
  further view-change.
\end{enumerate}

This cycle can repeat itself until the protocol reaches a period of synchrony;
at this point, view-changes will continue to occur until either the faulty
replicas allow the client's requests to complete, or a correct replica becomes
primary.
\end{proof}
\end{lemma}

\begin{theorem}[Liveness]
  All valid requests by a correct client will eventually complete.
  \begin{proof}
    We proceed by exhaustion.  Suppose the view is stable.  Then, by
    Lemma~\ref{liveness-sac}, a request will eventually complete.

    Now suppose the view is not stable. Then, by Lemma~\ref{liveness2-sac}, the
    view will eventually become stable. If the request completes before this
    occurs, then we are done. Otherwise, because the client retries its request
    continuously, the request will eventually arrive during a stable view, at
    which point by Lemma~\ref{liveness-sac} it will complete.
  \end{proof}
\end{theorem} }{
We include arguments for the safety and liveness of \name in
the extended version~\cite{saczyzzyva-arxiv}.
}
 \section{Performance evaluation}\label{sec:evaluation}
To assess the performance impact of our protocols, we created an experimental
setup, derived from~\cite{koen-thesis}, that runs proof-of-concept
implementations of Zyzzyva5 and \name
in a fault-free scenario.
Note that \name cannot be meaningfully compared
with regular Zyzzyva here, as they differ only in the presence of faults; we
might induce a fault ourselves, but in this case the performance of Zyzzyva is
mainly determined by the client timeout---that is, the time that the client waits
before broadcasting a commit certificate when it does not receive responses from
all replicas.  We therefore use Zyzzyva5 as a baseline for our experiments.

The trusted monotonic counter is implemented in an Intel SGX enclave, backed by
volatile memory.  The counter value is stored in an SGX-protected region of normal
memory, and set to zero at enclave initialization.  The use of volatile memory
provides high performance, and because a new counter instance is used for each view,
the loss of counter state in the event of a transient fault is not
catastrophic---if this happens, a view-change will occur but the protocol will continue.
All protocols are implemented using the same BFT platform, and so share
networking and cryptographic code.

We made our measurements using Amazon EC2~\cite{ec2} running a single replica
per instance, and a separate instance used by the client. Because EC2 does not
support SGX, the software was compiled in simulation
mode~\cite{sgx-sdk-developer-reference}. Separately, on a standalone SGX-enabled
machine, we confirmed that measurements in SGX simulation mode are similar to
measurements using SGX.

We report medians rather than mean and standard deviation, as the measured
latencies are non-normal.

\subsection{Performance within a single datacenter}\label{sec:performance-lan}

In order to test performance on low-latency networks, we carry out measurements
on a set of replicas placed within a single EC2 region, \emph{Frankfurt}. The test setup consists
of a cluster of 50 \texttt{m4.large} and \texttt{m5.large} EC2 instances~\cite{ec2}.

For each protocol, we measure the time it takes for a transaction to complete,
for increasing numbers of replicas, averaged over 50 transaction attempts.

\begin{figure}[!h]
\centering
\begin{subfigure}[b]{0.75\linewidth}
\centering
\includegraphics[width=0.7\linewidth]{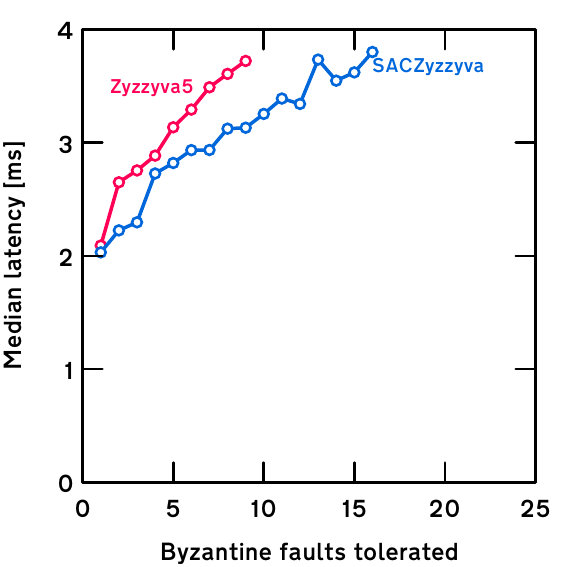}
  \caption{Latency vs tolerated faults within the \emph{Frankfurt} AWS region.}
  \label{fig:saczyzzyva-latency-lan}
\end{subfigure}
\vspace{2em}

\begin{subfigure}[b]{0.8\linewidth}
  \centering
    \includegraphics[width=0.7\linewidth]{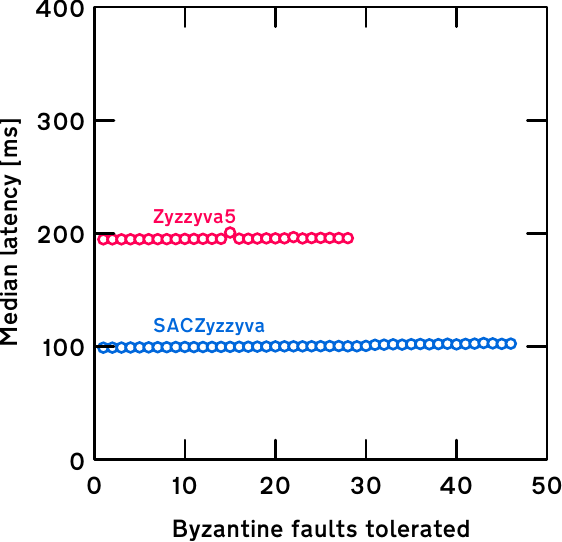}
  \caption{Latency vs tolerated faults across the internet.}
  \label{fig:saczyzzyva-latency-wan}
\end{subfigure}
 
\caption{Latency vs tolerated faults. Each latency is the median of 50 measurements. The number of
  tolerated faults $f$ is varied by modifying the number of replicas---$f$ faults are tolerated by
  $5f+1$ replicas for Zyzzyva5, and $3f+1$ replicas for 
  \name.}
\label{fig:saczyzzyva-latency}
\end{figure}

These results are shown in Figure~\ref{fig:saczyzzyva-latency-lan}. \name
requires fewer replicas than Zyzzyva5 for a given level of fault-tolerance, and
therefore completes requests in less time. While the number of replicas has a significant
effect on latency---on average, a marginal increase of 35$\mu$s/replica (\name) and
37$\mu$s/replica (Zyzzyva5)---the latency is still relatively small in an absolute
sense.  We will see in Section~\ref{sec:performance-wan} that latency is
dominated by network delays even with a larger number of replicas.

\subsection{Performance across the internet}\label{sec:performance-wan}
To assess the performance over high-latency networks such as the internet,
we measured the performance of \name and Zyzzyva5 using the replicas
divided between between three EC2 regions, \emph{Ohio}, \emph{Frankfurt}, and
\emph{Sydney} in order to approximate the performance of the protocols when
organically deployed across the internet.

In each test region we provision EC2 instances of type \texttt{m4.large} and
\texttt{m5.large}---50 in Frankfurt and Ohio, and 42
in Sydney, the maximum number available to us.

As in Section~\ref{sec:performance-lan}, we measure the response latency at the
client as a function of the number of tolerable faults. The results are shown in
Figure~\ref{fig:saczyzzyva-latency-wan}. Here latencies are dominated by speed-of-light delays, and increase
linearly at rates of 25$\mu$s/replica (SACZyzzyva) and 8$\mu$s/replica (Zyzzyva5) respectively.

In this particular geographic configuration, \name and significantly reduce
its latency by reducing the number of replies needed: Zyzzyva5 needs responses
from four fifths of replicas for requests to
complete, but \name requires only two thirds of replicas to respond.
This means that \name does not need to wait for responses to arrive across
the slow trans-Pacific link as Zyzzyva does. Another
surprising effect is that the rate of latency increase per replica is less than
when the protocol is run on a low-latency network. We hypothesize that this is
because the large network latencies mean that only the processing time of
responses from the most distant replicas affects the overall latency.

 \section{Optimality in the hybrid fault model}\label{sec:impossibility}

\begin{figure}
  \centering
  \includegraphics[width=0.9\linewidth]{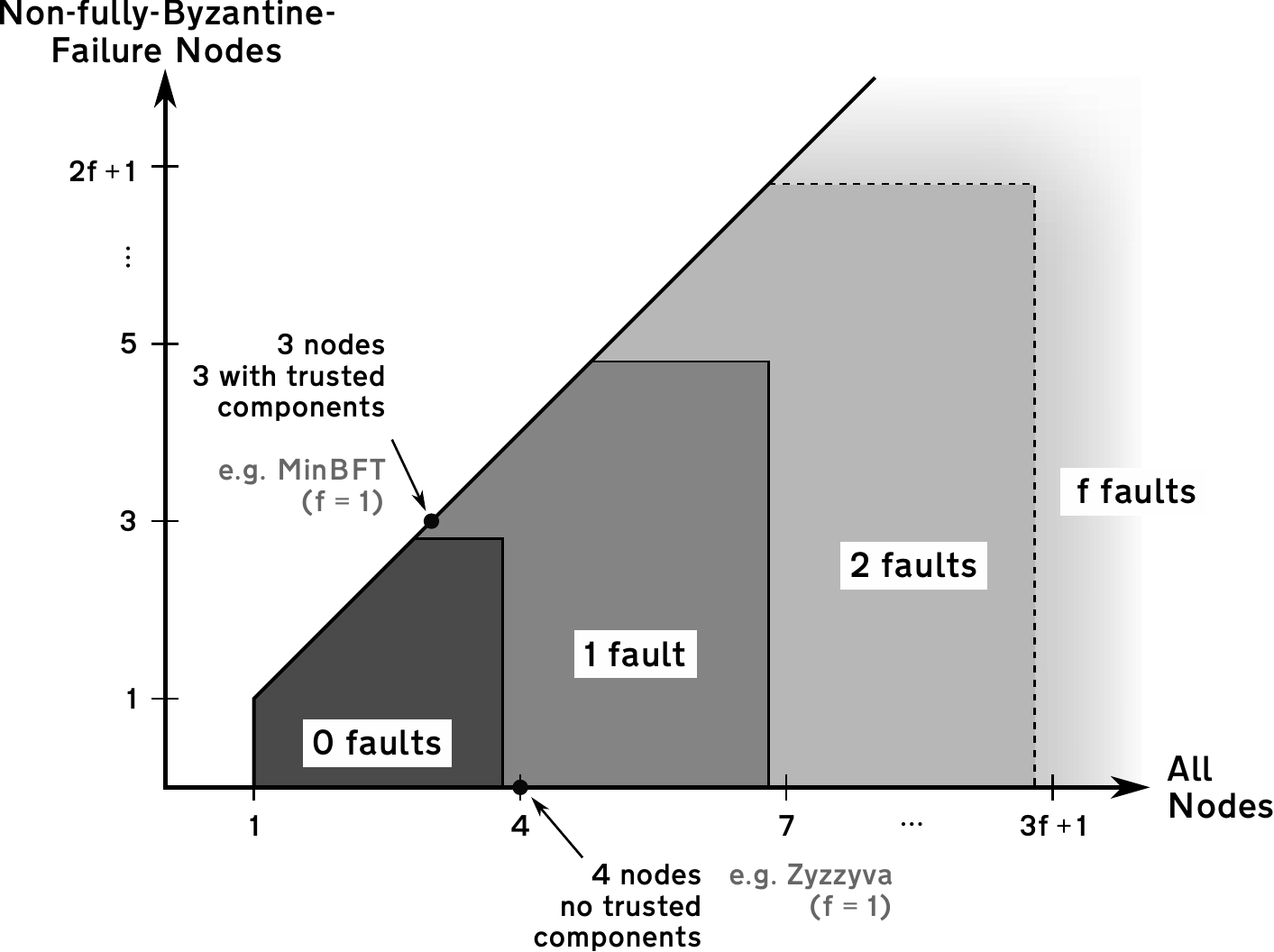}
  \caption{The level of fault tolerance achievable according to the total number
    of nodes and the number of nodes that cannot fail fully-Byzantine. Existing
    algorithms fall on the boundary of this space, for which the optimum
    fault tolerance is shown also in the interior.}
  \label{fig:design-space}
\end{figure}

Existing consensus protocols, to tolerate $f$ faults, require either $2f+1$
parties with a trusted component or $3f+1$ of any kind, as shown in
Figure~\ref{fig:design-space}; \name still requires $3f+1$
replicas despite the use of $f+1$ trusted monotonic counters, and it is reasonable to ask
whether these trusted components might allow us to obtain a similar protocol that
requires some smaller number of nodes.

We show here that this is not the case; specifically, that it is
impossible to achieve both safety and liveness without either $3f+1$ nodes in
total, or $2f+1$ nodes with trusted components. This theoretical limit is shown
graphically in Figure~\ref{fig:design-space}.

\subsection{Failure model}
We elaborate on the system model in Section~\ref{sec:system-model} by
introducing some new terminology.

\begin{figure}
  \centering
  \includegraphics[width=0.5\linewidth]{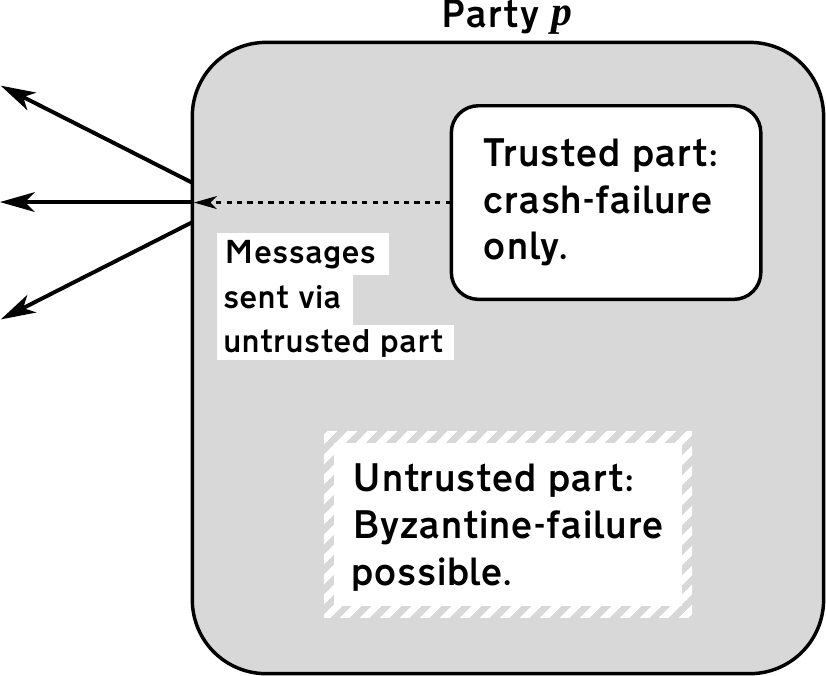}
  \caption{Hybrid model of trusted-component-equipped parties to the consensus protocol. Some
    parties will contain a trusted component that is immune from Byzantine
    failure: an attacker can make it crash or interfere with its communications,
    but cannot access its internal state.}
  \label{fig:hybrid}
\end{figure}

\textbf{Partially-Byzantine failures.} A party with a trusted
component can be split into two parts, as shown in
Figure~\ref{fig:hybrid}:
\begin{enumerate}
\item \emph{An untrusted part}, which either behaves correctly or suffers
  a Byzantine failure.
\item \emph{A trusted part}, which communicates via the untrusted
  part and either behaves correctly or suffers a crash failure.
\end{enumerate}

The result is that failures of a trusted-component-equipped party are
\emph{partially-Byzantine}: though their untrusted component can behave
arbitrarily, the trusted component will follow its programming, and thus other
parties can remain assured of at least some aspects of the behavior of the
party.

\textbf{Fully-Byzantine failures.}  We refer to the failures of a party without
a trusted component as
\emph{fully-Byzantine}: there are no restrictions on the behavior such a party
in the event of a failure.

\textbf{Crash failures.} In his failure mode, nodes simply crash. We refer to
crash and partially-Byzantine failures together as \textbf{non-fully-Byzantine}
failures.

More formally, we consider a set of parties $P$ executing a protocol $\pi$, and let
some subset $B$ be fully-Byzantine in the event of a failure, and its complement
$P \backslash B$ be `non-fully-Byzantine'.

We allow up to $f$ parties to fail according to their respective
failure modes: those failed parties that happen to be in in $B$ act under the full control of an adversary,
whereas those failed parties that are in $P \backslash B$ only give control of their \emph{untrusted parts}
to the adversary.

\iftoggle{impossibilityproof}{
  \subsection{Quorum properties}
We will proceed by a quorum-intersection argument, deriving some properties of
the quora of a consensus protocol, and then finding the conditions under which
they conflict. However, we must
re-examine this approach with the knowledge that some nodes may only be
\emph{partially}-Byzantine.  For the avoidance
of doubt, when we refer to an execution of a protocol by a set of parties, this
means that the \emph{correct} parties execute the protocol correctly, while
other parties can behave arbitrarily within the constraints of their failure
model.
\begin{definition}[Quorum]
  A set of parties $Q \subseteq P$ is a \emph{quorum} for a consensus protocol $\pi$ if,
  for any proposition $m$ by a proposer $p \in Q$, there exists some execution of
  $\pi$ by $Q$ in which no correct party receives any
  messages from parties $P \backslash Q$ outside $Q$, and some correct party $q \in Q$
  outputs $m$ after time at most $T(Q)$.
\end{definition}
Note that this definition does not require that the status of a quorum be a
determined by a simple threshold on the number of parties. In the case of PBFT,
any set of $2f+1$ parties is a quorum, but some protocol might conceivably give
greater weight to nodes with trusted components, or nodes that are known to have a lower
probability of failure.

A subtle point here is that for a set $Q$ to be a quorum, it is required only
that there \emph{exists} an execution of $\pi$ that leads to an output in time at
most $T(Q)$; for example, we might obtain some bound $T(Q)$ by simply observing
the consensus protocol in normal operation without introducing any adversarial
delays. This differs from the case of Byzantine quorum
systems~\cite{byzantine-quorum-systems}, where the set of quora is a design
parameter of the protocol.

The result is that, where the network model allows us to delay messages by time
$T(Q)$, we can delay messages between other nodes and some quorum, and there
will be some valid protocol execution that results in the correct parties
producing an output. We use this to show that the quora of any consensus
protocol with safety must have at least one non-fully-Byzantine node in their
intersection, mirroring the \emph{D-Consistency} property of a dissemination
quorum system in ~\cite[Definition 5.1]{byzantine-quorum-systems}.
\begin{lemma}[Quorum intersections cannot be fully-Byzantine]\label{lem:quorum-intersection}
  Let $Q_1$ and $Q_2$ be quora of a consensus protocol $\pi$ in the weak-synchrony
  model. Then, $Q_1 \cap Q_2$ contains at least one non-fully-Byzantine node.
  \begin{proof}
    By the safety of $\pi$, if any two correct parties to $\pi$ output values $m$
    and $m'$ respectively, then $m = m'$. We show that if $Q_1 \cap Q_2$ contains
    no non-fully-Byzantine nodes, then it is possible to force two correct
    parties to output distinct values.

    Let us define the sets $A = Q_1 \cap Q_2$, $B = Q_1 \backslash A$, and $C = Q_2 \backslash A$,
    and consider three possible runs:

    \textbf{Run 1.} Messages between $Q_1$ and $C$ are delayed for time
    $T(Q_1)$. Let some $p \in Q_1$ propose the value $m$. By the
    definition of a quorum, there is at least one protocol execution where a
    correct party in $Q_1$ outputs $m$.

    \textbf{Run 2.} Messages between $Q_2$ and $B$ are delayed for time
    $T(Q_2)$. Let some $p' \in Q_2$ propose the value $m'$. By the
    definition of a quorum, there is at least one protocol execution where a
    correct party in $Q_2$ outputs $m'$.

    \textbf{Run 3.} Now suppose messages between $B$ and $C$ are delayed for time
    $\max\{T(Q_1), T(Q_2)\}$. Let some $p \in B$ propose the value $m$, and some
    $p' \in C$ propose the value $m'$, $m \ne m'$. Suppose that $Q_1 \cap Q_2 = A$
    contains no non-fully-Byzantine nodes; then, we can have them behave arbitrarily.
    In this case, we have the nodes in $A$ behave as in Run~1 with respect to
    the nodes in $B$, and as in Run~2 with respect to the nodes in $C$. As the
    correct replicas in $B$ and $C$ cannot distinguish Run~3 from Runs~1 and~2
    respectively, then there is a protocol execution in which at least one
    correct node in each quorum will output the distinct values $m$ and $m'$
    respectively, thereby violating the assumption that the protocol is safe.

    Hence, $Q_1 \cap Q_2$ must contain at least one non-fully-Byzantine node.
  \end{proof}
\end{lemma}
The previous lemma gave a necessary---but not sufficient---condition for safety in
terms of quora.  Now, we do the same for liveness, mirroring the
\emph{D-Availability} property from~\cite[Definition~5.1]{byzantine-quorum-systems}.
\begin{lemma}[Sufficiently large sets must contain a quorum]\label{lem:quorum-existence}
  Let $S \subseteq P$ be a subset of parties to a consensus protocol $\pi$ tolerating $f$
  crash failures. Then, if $|S| \ge |P| - f$, $S$ is a quorum for $\pi$.
  \begin{proof}
    By the liveness of $\pi$, if a message $m$ is correctly proposed and the $|P|
    - |S| \le f$ parties $P \backslash S$ all crash, then all correct parties will
    eventually output some value. By the safety of $\pi$, the value that
    they output is $m$. Therefore, $H$ is a quorum.
  \end{proof}
\end{lemma}
 }
\subsection{Impossibility result}
\iftoggle{impossibilityproof}{
  With these two lemmas, we can now show our main result.
}{
  We describe here our main result; the proof appears in the extended version~\cite{saczyzzyva-arxiv}.
}
\begin{restatable}{theorem}{impossibility}\label{thm:impossibility}
  Let $\pi$ be a consensus protocol amongst $n$ parties in the partial synchrony
  model, $b$ of which, when they fail, fail fully-Byzantine, and $n-b$ of which,
  when they fail, either
  crash or fail partially-Byzantine. Then, to tolerate $f$ failures, at least one
  of the following must be true:
  \begin{align}
    n &\ge 3f + 1 \\
    n - b &\ge 2f + 1 .
  \end{align}

  \iftoggle{impossibilityproof}{  \begin{proof}
    We show that if neither condition holds,
    then if the protocol has liveness, it is not safe for at least one
    allocation of failures.

    Consider arbitrary $n \le 3f$ and $n-b \le 2f$. We proceed by contradiction.
    Suppose $\pi$ has liveness, and so Lemma~\ref{lem:quorum-existence} holds,
    Then, we seek some allocation of failures such that two quora $Q_1$ and
    $Q_2$ have only fully-Byzantine failures in their intersection.

    Let
    \begin{align*}
      Q_1 &= \left\{ 1, 2, \ldots, n - f \right\} \\
      Q_2 &= \left\{ f+1, f+2, \ldots, n \right\} .
    \end{align*}
    Because the numbering of the replicas is arbitrary, let us suppose that
    parties $B = \{\lfloor n/2 \rfloor - \lfloor b/2 \rfloor + 1, \ldots, \lfloor n/2 \rfloor + \lceil b/2 \rceil \}$ are subject
    to fully-Byzantine failure, as shown in
    Figure~\ref{fig:impossibility-quora}.
    \begin{figure}[h]
      \centering
      \includegraphics[width=0.65\linewidth]{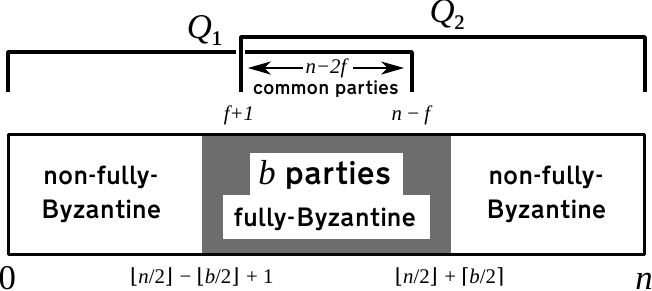}
      \caption{Constructed quora used in the proof of
        Theorem~\ref{thm:impossibility}, and the failure mode of each party.
        If the entire intersection can fail fully-Byzantine, then the protocol
        is unsafe.}
      \label{fig:impossibility-quora}
    \end{figure}

    Both $Q_1$ and $Q_2$ have cardinality
    $n-f$, so by Lemma~\ref{lem:quorum-existence}, they are both quora.

    Now, $|Q_1 \cap Q_2| = n-2f \le f$. Thus, we can make the entire intersection
    faulty. For $\pi$ to be safe---and thus Lemma~\ref{lem:quorum-intersection} to
    hold---this intersection must always contain at least one party that does not
    fail fully-Byzantine. But, this is not the case: $b \ge n - 2f$, hence
    \begin{align*}
      \min(B) &= \lfloor n/2 \rfloor - \lfloor b/2 \rfloor + 1 \\
              &\le \lfloor n/2 \rfloor - \lfloor n/2 - f \rfloor + 1\\
              &\le f+1 = \min Q_2 \\
      \end{align*}
      {and}
      \begin{align*}
      \max(B) &= \lfloor n/2 \rfloor + \lceil b/2 \rceil \\
                        &\ge \lfloor n/2 \rfloor + \lceil n/2 - f \rceil\\
                        &= n - f = \max Q_1 .
    \end{align*}
    Since $B = \{\min B, \ldots, \max B\}$, this implies $Q_1 \cap Q_2 \subseteq B$.

    We therefore have two quora that are not guaranteed to have a non-fully-Byzantine
    node in their intersection; this contradicts
    Lemma~\ref{lem:quorum-intersection}, and thus $\pi$ cannot have both liveness
    and safety if both $n \le 3f$ and $n-b \le 2f$.
  \end{proof}
 }{
    \begin{proof}
      The proof is given in the extended version~\cite{saczyzzyva-arxiv}.

      Briefly, we follow a quorum intersection argument. We define the concept
      of a quorum and prove that, for any quora $Q_1$ and $Q_2$, their
      intersection $Q_1 \cap Q_2$ must not fail fully-Byzantine. For this to be
      assured, $Q_1 \cap Q_2$ must contain at least one replica that cannot fail
      fully-Byzantine, or at least $f+1$ replicas in total, implying the
      conditions above.
    \end{proof}
  }
\end{restatable}
Therefore, it is impossible to outdo the usual requirement of $3f+1$ replicas
without $2f+1$ parties having access to some component that cannot fail
Byzantine. \section{Related Work}
As \name{} is motivated by recent blockchain-based distributed systems~\cite{bitcoin, ethereum},
in this section we review some research work that aims at scalability and efficiency for distributed consensus protocol involving large populations.

\noindent {\bf Consensus protocols in blockchain scenarios.}
Fabric~\cite{androulaki2018hyperledger} and Sawtooth~\cite{HLSawtooth} are two recent examples of distributed ledgers, which support the execution of \emph{smart contracts}.
Both use a consensus module to coordinate multiple parties.
In particular, Fabric can use a fault-tolerance protocol like such as
BFT-SMaRt~\cite{bessani2014state, sousa2017byzantine}, while Sawtooth is mostly
known for its Proof-of-Elapsed-Time protocol, which is vastly more scalable than
BFT protocols but
provides only eventual consistency. Protocols with $\mathcal{O}(n)$ message
complexity such as \name and CoSi allow for high scalability, as in Sawtooth,
but without sacrificing finality. 

Among other BFT protocols, there are also asynchronous protocols such as Honey
Badger~\cite{honeybadger} and BEAT~\cite{Duan2018}, which do not make any synchrony
assumptions.  However, this requires relatively expensive primitives such as
reliable broadcast and threshold cryptography, and so such protocols are less efficient.

Byzcoin~\cite{byzcoin} is a hybrid Nakamoto/BFT consensus protocol that uses the
Bitcoin consensus protocol to select a group of verifiers that is small enough
in size to run a traditional BFT algorithm. \name would serve well in this role,
as a replacement for the multisignature-based protocol used by~\cite{byzcoin}.

\noindent {\bf Protocols that reduce replica count.}
Several research works recognize the importance of tackling the equivocation
problem (malicious replicas sending out different conflicting messages to
different recipients) in BFT protocols, since this allows the reduction of the
replica count to $2f+1$.
MinBFT~\cite{minbft, correia2010asynchronous} proposes the use of a trusted
monotonic counter to tag the messages, making equivocation detectable.
Similarly, \cite{abraham2010fast} shows how to implement a weak sequenced
broadcast primitive using a TPM. \name{}'s use of trusted monotonic counters is closely related to MinBFT's approach.
A2M~\cite{Chun2007} provides an abstraction for attested append only memory. This is used to implement a hardware-based secure log for outgoing messages; while incoming messages are accepted only after the verification of a log attestation.
CheapBFT~\cite{Kapitza2012} and ReBFT~\cite{distler2016resource} provide a way to reduce further the number of replicas by making $f$ of them passive, and activating them only when it is required to handle faults and make progress.
\name{} puts a bridge between the world where \emph{all} replicas have a trusted component and the world where only \emph{some} of them have it, ultimately showing a protocol for the heterogeneous setting.

\noindent {\bf Protocols with low communication complexity.}
Several protocols have been proposed to reduce the message count.
Zyzzyva~\cite{Zyzzyva} and variants~\cite{guerraoui2010next} avoid all-to-all
broadcasts by using speculative execution.
Chain replication~\cite{van2004chain} has a low message complexity since replicas are organized on a chain-like communication topology and only use broadcasts in the case of faults.
Byzcoin~\cite{byzcoin} similarly uses a tree-like communication topology, and
uses collective signing to aggregate messages from multiple nodes.
FastBFT~\cite{fastbft} improves on that approach by means of a lightweight TEE-based message aggregation technique.
\name{} belongs to the former category, using speculative execution to reduce
the number of messages, but without needing to make the trade-off between
fault-tolerance and robust performance as with Zyzzyva.

\noindent {\bf Lower bounds.}
BFT protocols suffer from several fundamental limitations.
First, it has been shown~\cite{lamport2006lower} that asynchrnous protocols
require two phases to terminate. Speculative protocols like Zyzzyva or \name{}
are able to terminate in one phase since they make additional assumption
(namely, that rollback is possible).
Second, BFT protocols typically require a quadratic number of messages to terminate~\cite{DolevReischuk}. The workaround for many protocols is to use cryptographic constructions which can err with positive probability~\cite{king2011breaking}.
Finally, in~\cite{clement2012limited} it has been shown that achieving
\emph{non-equivocation} is actually insufficient for reducing the number of
replicas, and that \emph{transferable authentication} of messages (e.g., using
digital signatures) is additionally necessary. In \name{} the trusted monotonic
counter ensures non-equivocation, with an attestation that is publicly
verifiable (and so transferable) with the aid of digital certificates from the hardware manufacturer. \section{Discussion and Conclusions}
By incorporating a trusted monotonic counter into Zyzzyva's
ordering process, we can eliminate its non-speculative fallback without
sacrificing fault-tolerance as previous solutions have. This removes one of the
main disadvantages of the Zyzzyva family of protocols, namely that without
sacrificing fault-tolerance they are unable to perform speculative execution in
the presence of even a single fault.

\name achieves the resilience of Zyzzyva5 while reducing the replica count from
$5f+1$ to $3f+1$. MinZyzzyva uses trusted monotonic counters in every replica,
and so in principle we might expect that MinZyzzyva's non-speculative fallback
can be similarly eliminated. This is not entirely straightforward, as we need to
ensure that even a faulty replica will disclose the requests that it has seen.
We will address this topic later in an extended version of this paper.

Our approach does not only apply to Zyzzyva-like
protocols. For example, PBFT uses an all-to-all broadcast to provide a
canonical ordering of requests; when a trusted monotonic counter is available,
this step can be eliminated, as in MinBFT~\cite[Figure~1]{minbft}, but without
requiring a trusted monotonic counter in every replica~\cite{koen-thesis}.

We have also shown that more than two-thirds of replicas must have a trusted
component in order to tolerate more than $\lfloor (n-1)/3 \rfloor$ faults. This means that
our protocols achieve optimal fault-tolerance, but shows that there is an
important part of the design space that remains unexplored.
 \section{Acknowledgements}
This work is supported in part by the Academy of Finland (grant~309195) and by Intel (ICRI-CARS). \bibliographystyle{plain}
\bibliography{references}

\begin{thebibliography}{10}

\bibitem{ec2}
Amazon {EC2}.
\newblock https://aws.amazon.com/ec2/.

\bibitem{trustzone}
{ARM} security technology: Building a secure system using {TrustZone}
  technology.
\newblock White paper, ARM, 2009.

\bibitem{sgx-sdk-developer-reference}
Intel {Software} {Guard} {Extensions} {SDK} for {Linux} {OS}: Developer
  reference.
\newblock Technical report, 2016.

\bibitem{abraham2010fast}
Ittai Abraham, Marcos~K Aguilera, and Dahlia Malkhi.
\newblock Fast asynchronous consensus with optimal resilience.
\newblock In {\em International Symposium on Distributed Computing}, pages
  4--19. Springer, 2010.

\bibitem{abraham2018revisiting}
Ittai Abraham, Guy Gueta, Dahlia Malkhi, and Jean-Philippe Martin.
\newblock Revisiting fast practical {Byzantine} fault tolerance: {Thelma},
  {Velma}, and {Zelma}.
\newblock {\em arXiv preprint arXiv:1801.10022}, 2018.

\bibitem{androulaki2018hyperledger}
Elli Androulaki, Artem Barger, Vita Bortnikov, Christian Cachin, Konstantinos
  Christidis, Angelo De~Caro, David Enyeart, Christopher Ferris, Gennady
  Laventman, Yacov Manevich, et~al.
\newblock Hyperledger fabric: a distributed operating system for permissioned
  blockchains.
\newblock In {\em Proceedings of the Thirteenth EuroSys Conference}, page~30.
  ACM, 2018.

\bibitem{barborak1993consensus}
Michael Barborak, Anton Dahbura, and Miroslaw Malek.
\newblock The consensus problem in fault-tolerant computing.
\newblock {\em ACM Computing Surveys}, 25(2):171--220, 1993.

\bibitem{bessani2014state}
Alysson Bessani, Jo{\~a}o Sousa, and Eduardo~EP Alchieri.
\newblock State machine replication for the masses with {BFT-SMART}.
\newblock In {\em Dependable Systems and Networks (DSN), 2014 44th Annual
  IEEE/IFIP International Conference on}, pages 355--362. IEEE, 2014.

\bibitem{bonneau2015sok}
Joseph Bonneau, Andrew Miller, Jeremy Clark, Arvind Narayanan, Joshua~A Kroll,
  and Edward~W Felten.
\newblock Sok: Research perspectives and challenges for {Bitcoin} and
  cryptocurrencies.
\newblock In {\em Security and Privacy (SP), 2015 IEEE Symposium on}, pages
  104--121. IEEE, 2015.

\bibitem{ethereum}
Vitalik. Buterin.
\newblock A next-generation smart contract and decentralized application
  platform, 2014.
\newblock \url{https://github.com/ethereum/wiki/wiki/White-Paper}.

\bibitem{pbft}
Miguel Castro and Barbara Liskov.
\newblock Practical {Byzantine} fault tolerance.
\newblock In {\em Proceedings of the Third Symposium on Operating Systems
  Design and Implementation}, OSDI '99, pages 173--186, Berkeley, CA, USA,
  1999. USENIX Association.

\bibitem{Chun2007}
Byung-Gon Chun, Petros Maniatis, Scott Shenker, and John Kubiatowicz.
\newblock Attested append-only memory: Making adversaries stick to their word.
\newblock In {\em Proceedings of Twenty-first ACM SIGOPS Symposium on Operating
  Systems Principles}, SOSP '07, pages 189--204, 2007.

\bibitem{clement2012limited}
Allen Clement, Flavio Junqueira, Aniket Kate, and Rodrigo Rodrigues.
\newblock On the (limited) power of non-equivocation.
\newblock In {\em Proceedings of the 2012 ACM symposium on Principles of
  distributed computing}, pages 301--308. ACM, 2012.

\bibitem{CWADM09}
Allen Clement, Edmund Wong, Lorenzo Alvisi, Mike Dahlin, and Mirco Marchetti.
\newblock Making {Byzantine} fault tolerant systems tolerate {Byzantine}
  faults.
\newblock In {\em Proceedings of the 6th USENIX Symposium on Networked Systems
  Design and Implementation}, pages 153--168, 2009.

\bibitem{correia2010asynchronous}
Miguel Correia, Giuliana~S Veronese, and Lau~Cheuk Lung.
\newblock Asynchronous {Byzantine} consensus with $2f+ 1$ processes.
\newblock In {\em Proceedings of the 2010 ACM symposium on applied computing},
  pages 475--480. ACM, 2010.

\bibitem{distler2016resource}
Tobias Distler, Christian Cachin, and R{\"u}diger Kapitza.
\newblock Resource-efficient {Byzantine} fault tolerance.
\newblock {\em IEEE Transactions on Computers}, 65(9):2807--2819, 2016.

\bibitem{DolevReischuk}
Danny Dolev and R\"{u}diger Reischuk.
\newblock Bounds on information exchange for {Byzantine} agreement.
\newblock {\em J. ACM}, 32(1):191--204, January 1985.

\bibitem{Duan2018}
Sisi Duan, Michael~K. Reiter, and Haibin Zhang.
\newblock {BEAT}: Asynchronous {BFT} made practical.
\newblock In {\em Proceedings of the 2018 ACM SIGSAC Conference on Computer and
  Communications Security}, CCS '18, pages 2028--2041, New York, NY, USA, 2018.
  ACM.

\bibitem{fischer1985impossibility}
Michael~J Fischer, Nancy~A Lynch, and Michael~S Paterson.
\newblock Impossibility of distributed consensus with one faulty process.
\newblock {\em J. ACM}, 32(2):374--382, 1985.

\bibitem{TCGTPM}
Trusted~Computing Group.
\newblock {Trusted Platform Module specification}, 2016.
\newblock
  \url{https://trustedcomputinggroup.org/resource/tpm-library-specification/}.

\bibitem{guerraoui2010next}
Rachid Guerraoui, Nikola Kne{\v{z}}evi{\'c}, Vivien Qu{\'e}ma, and Marko
  Vukoli{\'c}.
\newblock The next 700 {BFT} protocols.
\newblock In {\em Proceedings of the 5th European conference on Computer
  systems}, pages 363--376. ACM, 2010.

\bibitem{gueta2018sbft}
Guy~Golan Gueta, Ittai Abraham, Shelly Grossman, Dahlia Malkhi, Benny Pinkas,
  Michael~K Reiter, Dragos-Adrian Seredinschi, Orr Tamir, and Alin Tomescu.
\newblock {SBFT}: a scalable decentralized trust infrastructure for
  blockchains.
\newblock {\em arXiv preprint arXiv:1804.01626}, 2018.

\bibitem{HLSawtooth}
Hyperledger.
\newblock Sawtooth.
\newblock \url{www.hyperledger.org/projects/sawtooth}.

\bibitem{sgx}
Intel.
\newblock {Software Guard Extensions (Intel SGX) Programming Reference}, 2013.
\newblock
  \url{https://software.intel.com/sites/default/files/managed/48/88/329298-002.pdf}.

\bibitem{Kapitza2012}
R\"{u}diger Kapitza, Johannes Behl, Christian Cachin, Tobias Distler, Simon
  Kuhnle, Seyed~Vahid Mohammadi, Wolfgang Schr\"{o}der-Preikschat, and Klaus
  Stengel.
\newblock {CheapBFT}: Resource-efficient {Byzantine} fault tolerance.
\newblock In {\em Proceedings of the 7th ACM European Conference on Computer
  Systems}, EuroSys '12, pages 295--308, New York, NY, USA, 2012. ACM.

\bibitem{king2011breaking}
Valerie King and Jared Saia.
\newblock Breaking the ${O}(n^2)$ bit barrier: scalable {Byzantine} agreement
  with an adaptive adversary.
\newblock {\em Journal of the ACM (JACM)}, 58(4):18, 2011.

\bibitem{byzcoin}
Eleftherios~Kokoris Kogias, Philipp Jovanovic, Nicolas Gailly, Ismail Khoffi,
  Linus Gasser, and Bryan Ford.
\newblock Enhancing {Bitcoin} security and performance with strong consistency
  via collective signing.
\newblock In {\em 25th {USENIX} Security Symposium ({USENIX} Security 16)},
  pages 279--296, Austin, TX, 2016. {USENIX} Association.

\bibitem{Zyzzyva}
Ramakrishna Kotla, Lorenzo Alvisi, Mike Dahlin, Allen Clement, and Edmund Wong.
\newblock Zyzzyva: Speculative {Byzantine} fault tolerance.
\newblock In {\em Proceedings of Twenty-first ACM SIGOPS Symposium on Operating
  Systems Principles}, SOSP '07, pages 45--58, New York, NY, USA, 2007. ACM.

\bibitem{lamport2006lower}
Leslie Lamport.
\newblock Lower bounds for asynchronous consensus.
\newblock {\em Distributed Computing}, 19(2):104--125, 2006.

\bibitem{lamport1982byzantine}
Leslie Lamport, Robert Shostak, and Marshall Pease.
\newblock The {Byzantine} generals problem.
\newblock {\em ACM Transactions on Programming Languages and Systems (TOPLAS)},
  4(3):382--401, 1982.

\bibitem{levin2009trinc}
Dave Levin, John~R Douceur, Jacob~R Lorch, and Thomas Moscibroda.
\newblock {TrInc}: Small trusted hardware for large distributed systems.
\newblock In {\em Proceedings of NSDI}, volume~9, pages 1--14, 2009.
\newblock Boston, MA, USA.

\bibitem{fastbft}
Jian Liu, Wenting Li, Ghassan~O Karame, and N.~Asokan.
\newblock Scalable {Byzantine} consensus via hardware-assisted secret sharing.
\newblock {\em IEEE Transactions on Computers}, 2018.

\bibitem{lynch-distributed-algorithms}
Nancy~A Lynch.
\newblock {\em Distributed Algorithms}.
\newblock Morgan Kaufmann, 1996.

\bibitem{malkhi2018blockchain}
Dahlia Malkhi.
\newblock Blockchain in the lens of {BFT}.
\newblock In {\em USENIX Annual Technical Conference}, 2018.
\newblock Boston, MA, USA.

\bibitem{byzantine-quorum-systems}
Dahlia Malkhi and Michael Reiter.
\newblock Byantine quorum systems.
\newblock {\em Distributed Computing}, 11(4):203--213, 1998.

\bibitem{martin2005fast}
J-P Martin and L~Alvisi.
\newblock Fast {Byzantine} consensus.
\newblock In {\em Proceedings of the International Conference on Dependable
  Systems and Networks (DSN)}, pages 402--411. IEEE, 2005.

\bibitem{matetic2017rote}
Sinisa Matetic, Mansoor Ahmed, Kari Kostiainen, Aritra Dhar, David Sommer,
  Arthur Gervais, Ari Juels, and Srdjan Capkun.
\newblock {ROTE}: Rollback protection for trusted execution.
\newblock {\em IACR Cryptology ePrint Archive}, 2017:48, 2017.

\bibitem{honeybadger}
Andrew Miller, Yu~Xia, Kyle Croman, Elaine Shi, and Dawn Song.
\newblock The honey badger of {BFT} protocols.
\newblock In {\em Proceedings of the 23rd ACM Conference on Computer and
  Communications Security (CCS '16)}, 2016.

\bibitem{bitcoin}
Satoshi Nakamoto.
\newblock Bitcoin: A peer-to-peer electronic cash system, 2009.
\newblock \url{http://www.bitcoin.org/bitcoin.pdf}.

\bibitem{minbft}
Giuliana Santos~Veronese, Miguel Correia, Alysson Neves~Bessani, Lau~Cheuk
  Lung, and Paulo Verissimo.
\newblock Efficient {Byzantine} fault-tolerance.
\newblock {\em IEEE Transactions on Computers}, 62:16--30, 2013.

\bibitem{sousa2017byzantine}
Joao Sousa, Alysson Bessani, and Marko Vukoli{\'c}.
\newblock A {Byzantine} fault-tolerant ordering service for the {Hyperledger}
  {Fabric} blockchain platform.
\newblock {\em arXiv:1709.06921}, 2017.

\bibitem{strackx2016ariadne}
Raoul Strackx and Frank Piessens.
\newblock Ariadne: A minimal approach to state continuity.
\newblock In {\em USENIX Security}, volume~16, 2016.
\newblock Austin, TX, USA.

\bibitem{cosi}
Ewa Syta, Iulia Tamas, Dylan Visher, David~Isaac Wolinsky, Philipp Jovanovic,
  Linus Gasser, Nicolas Gailly, Khoffi, Ismail, and Bryan Ford.
\newblock Keeping authorities ``honest or bust'' with decentralized witness
  cosigning.
\newblock In {\em 37th IEEE Symposium on Security and Privacy}, 2016.

\bibitem{koen-thesis}
Koen Tange.
\newblock High speed consensus with trusted execution environments.
\newblock Master's thesis, Aalto University, 2018.

\bibitem{van2004chain}
Robbert Van~Renesse and Fred~B Schneider.
\newblock Chain replication for supporting high throughput and availability.
\newblock In {\em OSDI}, volume~4, pages 91--104, 2004.

\bibitem{hybridization}
Paulo Ver\'{\i}ssimo.
\newblock Travelling through wormholes: a new look at distributed systems
  models.
\newblock {\em ACM SIGACT News}, 37(1):66--81, 2006.

\bibitem{vukolic2015quest}
Marko Vukoli{\'c}.
\newblock The quest for scalable blockchain fabric: Proof-of-work vs. {BFT}
  replication.
\newblock In {\em International Workshop on Open Problems in Network Security},
  pages 112--125. Springer, 2015.

\bibitem{vukolic2016eventually}
Marko Vukolic.
\newblock Eventually returning to strong consistency.
\newblock {\em IEEE Data Eng. Bull.}, 39(1):39--44, 2016.

\end{thebibliography}

\end{document}